\documentclass[useAMS,usenatbib]{mn2e}

\usepackage{aas_macros}
\usepackage{graphicx}
\usepackage{amsmath}
\usepackage{xspace}
\usepackage{ifthen}
\usepackage{lscape}
\usepackage{placeins}
\usepackage{longtable}
\usepackage{hyperref}

\usepackage{array}
\newcolumntype{L}[1]{>{\raggedright\let\newline\\\arraybackslash\hspace{0pt}}p{#1}}
\newcolumntype{C}[1]{>{\centering\let\newline\\\arraybackslash\hspace{0pt}}p{#1}}
\newcolumntype{R}[1]{>{\raggedleft\let\newline\\\arraybackslash\hspace{0pt}}p{#1}}

\newcommand{\teff}{$T_\text{eff}$\xspace}    
\newcommand{\logg}{$\log g$\xspace}

\title[Doppler tomography with TRES]{Spin orbit alignment for KELT-7b and HAT-P-56b via Doppler tomography with TRES}

\author[G.~Zhou et al.]
{\parbox{\textwidth}
  {George Zhou$^{1}$\thanks{E-mail: \texttt{george.zhou@cfa.harvard.edu}},
    David W. Latham$^{1}$,
    Allyson Bieryla$^{1}$,
    Thomas G. Beatty$^{2,3}$,
    Lars A. Buchhave$^{4,1}$,
    Gilbert A. Esquerdo$^{1}$,
    Perry Berlind$^{1}$,
    and Michael L. Calkins$^{1}$
\vspace{0.4cm}}\\
\parbox{\textwidth}{
  $^{1}${Harvard-Smithsonian Center for Astrophysics, 60 Garden St., Cambridge, MA 02138, USA}\\
  $^{2}${Department of Astronomy \& Astrophysics, The Pennsylvania State University, 525 Davey Lab, University Park, PA 16802, USA}\\
  $^{3}${Center for Exoplanets and Habitable Worlds, The Pennsylvania State University, 525 Davey Lab, University Park, PA 16802, USA}\\
  $^{4}${Centre for Star and Planet Formation, Natural History Museum of Denmark, University of Copenhagen, DK-1350 Copenhagen, Denmark}\\
}}
\begin{document}
%\linenumbers

\date{Accepted 2016-05-06}

\pagerange{\pageref{firstpage}--\pageref{lastpage}} \pubyear{2016}

\maketitle

\label{firstpage}

\begin{abstract}
We present Doppler tomographic analyses for the spectroscopic transits of KELT-7b and HAT-P-56b, two hot-Jupiters orbiting rapidly rotating F-dwarf host stars. These include analyses of archival TRES observations for KELT-7b, and a new TRES transit observation of HAT-P-56b. We report spin-orbit aligned geometries for KELT-7b ($2.7 \pm 0.6\,^\circ$) and HAT-P-56b ($8 \pm 2\,^\circ$). The host stars KELT-7 and HAT-P-56 are among some of the most rapidly rotating planet-hosting stars known. We examine the tidal re-alignment model for the evolution of the spin-orbit angle in the context of the spin rates of these stars. We find no evidence that the rotation rates of KELT-7 and HAT-P-56 have been modified by star-planet tidal interactions, suggesting that the spin-orbit angle of systems around these hot stars may represent their primordial configuration. In fact, KELT-7 and HAT-P-56 are two of three systems in super-synchronous, spin-orbit aligned states, where the rotation periods of the host stars are faster than the orbital periods of the planets.
\end{abstract}


\section{Introduction}
\label{sec:introduction}

The observed population of hot-Jupiters is thought to have migrated inward after their formation. The angle between the spin axis of the star and the orbit normal of the hot-Jupiter is a useful probe for the migration history of the planet. In the most simple interpretation, planets found in well-aligned orbits are thought to have migrated in the protoplanetary disk via planet-gas interactions \citep[e.g.][]{1996Natur.380..606L}, while those found in high obliquity orbits underwent dynamical interactions, such as planet-planet scattering \citep[e.g.][]{1996Sci...274..954R}, Kozai-Lidov induced eccentricity migration \citep[e.g.][]{2003ApJ...589..605W,2007ApJ...669.1298F}, or were born in primordially tilted disks \citep[e.g.][]{2010MNRAS.401.1505B,2012Natur.491..418B}.  Of the 74 planets with  spin-orbit measurements\footnote{Measured by the Rossiter-McLaughlin effect, a technique that does not impose strong selection biases on the spin-orbit orientation of the systems measured. Sample selected from R\'{e}ne Heller's Holt-Rossiter-McLaughlin Encyclopedia (\url{http://www2.mps.mpg.de/homes/heller/}). Where multiple spin-orbit angles are quoted, the authors examined the discovery paper and chose the most robust observation. Only hot-Jupiters are included.}, 23\% are found in misaligned orbits.

However, interpreting the misalignment statistic is made harder by potential post-migration evolution of the orbit geometry. It is thought that star-planet interactions, such as tidal and magnetic drag, can realign the spin direction of the convective envelope of a host star \citep{2010ApJ...718L.145W,2012MNRAS.423..486L,2013ApJ...769L..10R,2014ApJ...784...66X,2014ApJ...790L..31D}. This is supported by the observed trend that massive planets in close-in orbits around cooler stars tend to be aligned, while smaller planets around hotter stars (which lack convective envelopes), or at longer periods (where tidal forces are weak), exhibit a wide range of obliquity angles \citep[e.g.][]{2010ApJ...718L.145W,2010ApJ...719..602S,2012ApJ...757...18A,2015ApJ...801....3M}.  It should be noted that this framework has some observational shortcomings. \citet{2015ApJ...801....3M} and \citet{2015arXiv151105570L} compared the photometric variability (a potential proxy for line-of-sight spin axis inclination) of planet-hosting stars to other stars of similar properties that do not host transiting hot-Jupiters. They found the stellar type -- spin-orbit angle trend persists at long periods, beyond the bounds of tidal interactions. In addition, short period planets around cool stars have been found in severe misalignment \citep{2010MNRAS.402L...1P,2015ApJ...814L..16Z}, these planets should have realigned the convective envelope of their stars under the tidal theory.

Within this tidal realignment framework, we can postulate that the primordial spin-orbit angles of planets around early-type stars are more likely recoverable, especially those that exhibit rapid rotation and have not yet been spun-down by planet-star interactions. Unlike late-type stars that undergo magnetic braking, stars hotter than $T_\text{eff} \sim 6250\,\text{K}$ do not spin-down significantly, and are generally more rapidly rotating. One key problem with characterising planets around rapidly rotating, early type stars is that precise radial velocity measurements needed for the Rossiter-McLaughlin effect \citep{1924ApJ....60...15R,1924ApJ....60...22M} are difficult to obtain.

In this study, we present Doppler tomographic analyses to measure the spin-orbit angles of two systems, orbiting rapidly rotating F-type stars. We present a new spectroscopic transit observation for the hot-Jupiter HAT-P-56b \citep{2015AJ....150...85H}, and a re-analysis of archival observations for the hot-Jupiter KELT-7b \citep{2015AJ....150...12B}. The key properties of these systems are presented in Table~\ref{tab:literature_properties}. They all orbit rapidly rotating F-dwarfs, with projected rotational velocities of $70$, and $38\,\text{km\,s}^{-1}$, respectively (see Section~\ref{sec:lsd}). Of all transiting hot-Jupiter hosts known, only the A-stars WASP-33 \citep{2010MNRAS.407..507C}, KOI-13 \citep{2014ApJ...790...30J}, and HAT-P-57 \citep{2015AJ....150..197H} have higher rotation rates.

\begin{table*}
  
  \caption{\label{tab:literature_properties}Key properties of the KELT-7 and HAT-P-56 systems from the literature}
  \begin{tabular}{lrr}
    \hline\hline
    &  \textbf{KELT-7} & \textbf{HAT-P-56}\\
    Source & \citet{2015AJ....150...12B} & \citet{2015AJ....150...85H}\\
    \hline
    RA &  05:13:11.0 & 06:45:24.0\\
    DEC &  +33:19:05 & +27:15:08\\
    $V_\text{mag}$ &  8.54 & 10.91\\
    $M_\star \, (M_\odot)$ &  $1.535_{-0.054}^{+0.066}$ & $1.296\pm0.036$ \\
    $R_\star \, (R_\odot)$ &  $1.732_{-0.045}^{+0.043}$ & $1.428\pm0.030$\\
    $T_\text{eff}\,(\text{K})$  &$6789_{-49}^{+50}$ & $6566\pm50$\\
    $v\sin i\,(\text{km\,s}^{-1})$  & $65.0_{-5.9}^{+6.0}$ & $40.06\pm0.50$\\
    $M_p\,(M_\text{Jup})$  & $1.28\pm0.18$ & $2.18\pm0.25$\\
    $R_p\,(R_\text{Jup})$  & $1.533_{-0.047}^{+0.046}$ & $1.466\pm0.040$\\
    Period (days)  & $2.7347749\pm0.0000039$ & $2.7908327\pm0.0000047$\\
    $|\lambda|\,(^\circ)$  & $9.7\pm5.2$ & -\\
    \hline
  \end{tabular}
\end{table*}

The vast majority of known spin-orbit angles have been measured via the Rossiter-McLaughlin effect. When the planet transits the host star, it successively blocks out part of the rotating stellar surface, and thereby induces a net shift in the centroid of the stellar spectral lines, measured as an apparent in-transit radial velocity variation that is dependent on the transit geometry. The Rossiter-McLaughlin effect was observed for KELT-7b by \citet{2015AJ....150...12B}. The system was reported to be in prograde, spin-orbit aligned geometries. The spectral lines of rapidly rotating host stars are severely broadened by rotation, resulting in blending of individual lines. In such cases it is often possible to directly measure the deviation in the rotational broadening kernel of the spectral lines induced by the transiting planet -- a technique known as Doppler tomography. The technique has already been employed to measure the spin-orbit angles of a number of planetary systems \citep{2010MNRAS.403..151C,2010MNRAS.407..507C,2010A&amp;A...523A..52M,2012ApJ...760..139B,2012A&amp;A...543L...5G,2013ApJ...771...11A,2014ApJ...790...30J,2015ApJ...810L..23J,2015AJ....150..197H}. This technique allows us to directly detect the Doppler shadow of the planet, providing a more accurate measurement of the spin-orbit angle, as well improved characterisation of other transit parameters. We present spin-orbit angles for KELT-7b and HAT-P-56b measured via the Doppler tomographic technique.

\section{Spectroscopic observations and reductions}
\label{sec:obs_and_analyses}

\subsection{TRES transit spectroscopy observations}
\label{sec:tres}

Spectroscopic observations of the transits were obtained with the Tillinghast Reflector Echelle Spectrograph (TRES) on the 1.5\,m telescope at the Fred Lawrence Whipple Observatory, Mount Hopkins, Arizona, USA. The spectrograph has a resolving power of $\lambda / \Delta \lambda \equiv R = 44000$, sampling the spectral range of 3850-9100\,\AA\, over 51 echelle orders. We use the archival spectroscopic transit data of KELT-7b, originally presented by \citet{2015AJ....150...12B}. The observation was performed on 2013-10-19 UT, a total of 29 spectra were observed, each with 900\,s exposure time made up of three exposures in order to optimize the removal of cosmic rays. The set of spectra achieved an average signal-to-noise per resolution element of $\text{S/N} \sim 160$ over the Mg b line region. Wavelength calibration is achieved by a sequence of Th-Ar hollow cathode lamp exposures that bracket each 900s exposure. The transit of HAT-P-56b was observed on 2016-01-03 UT, with a total of 30 spectra obtained. Similar to the KELT-7b observations, each spectrum was combined from an average of three short exposures, with a total exposure time of 540\,s, achieving an average S/N of $\sim35$. Details of the spectral reduction and extraction is similar to that described in \citet{2010ApJ...720.1118B} for FIES observations. The pipeline was modified to work with the TRES spectrograph. All the parameters pertaining to the detector and spectrograph were modified, like the CCD format including overscan regions, the gain and readout noise, an initial guess for the position of the ThAr lines. Furthermore, tweaks were made for the handling of 3D cosmic ray removal, 3D profiling to remove pixel-to-pixel variations. However, these modifications are mostly minor changes and the bulk of the pipeline did not need major modifications to work with TRES.

\subsection{Retrieving the stellar broadening profile}
\label{sec:lsd}

Extracting the line profile from the spectra of rapidly rotating stars is complicated by the lack of unblended lines. We follow the technique set out in \citet{1997MNRAS.291..658D} and \citet{2010MNRAS.407..507C}, and perform a Least-Squares Deconvolution (LSD) to recover the line broadening kernel of each spectrum. This usually involves deconvolving the observed spectrum against a weighted delta-function line list to derive the broadening kernel of the star. Following \citet{2015AJ....150..197H}, we use unbroadened synthetic spectral templates, rather than weighted delta functions, as the template of the deconvolution. Synthetic spectral templates are generated using the spectral synthesis program SPECTRUM\footnote{\texttt{http://www1.appstate.edu/dept/physics/spectrum/spectrum.html}} \citep{1994AJ....107..742G}, with the ATLAS9 model atmospheres \citep{2004astro.ph..5087C}. We assume no rotation, microturbulence, macroturbulence, and instrumental broadening for the spectral template. 

Each echelle order of the TRES spectrum is first blaze corrected and continuum normalised. We then derive broadening profiles for consecutive sections of the spectrum, each spanning three echelle orders $(\sim 200\,\text{\AA})$. A 20\% trapezium apodisation is applied to the observed spectrum and template to reduce the artifacts that are induced by the deconvolution. The broadening profiles from each section are average combined to form the final rotational profile for each exposure. A total of 34 echelle orders were used, spanning the spectral range 3900--6250\,\AA. This region was chosen to best avoid the telluric absorption lines. We found that deconvolutions of spectra stitched from three consecutive echelle orders yielded lower noise in the final rotational profile than either deconvolution of individual echelle orders, or deconvolution of the entire stitched spectrum. The $\sim 200\,$\AA\, long spectral regions contain enough information to allow an effective deconvolution, and are small enough that the sections can be weighted to arrive at the highest signal-to-noise averaged profile. The radial velocity shift of the star through the transit sequence, determined from the published orbit, is then subtracted, such that the centroid of each rotation kernel is shifted to $0\,\text{km\,s}^{-1}$.

\subsection{Measuring $v\sin i$ from broadening profile}
\label{sec:vsini}
Spin-orbit angles derived from Doppler tomography and Rossiter-McLaughlin analyses are often degenerate with the rotational velocity of the star. However, $v \sin i$ is difficult to measure from the spectrum due to degenerate effects with other broadening parameters, such as macro turbulence, and the assumed limb darkening parameters. \citet{2012ApJ...757..161T} found the $v\sin i$ estimates from the Stellar Parameter Classification (SPC) pipeline \citep[used in the discovery papers,] []{2012Natur.486..375B} were systematic offset from those of the Spectroscopy Made Easy \citep[SME][]{1996A&amp;AS..118..595V} analyses. Additional systematic offsets in \teff and \logg between the spectral retrieval procedures have been noted \cite[e.g.][]{2012ApJ...757..161T,2013A&A...558A.106M}, but are small enough that the derived line profiles and Doppler tomographic signals are not affected. However, we note that the host star properties, which are part of the global modelling in Section~\ref{sec:modelling}, will affect the final derived stellar properties (e.g. $M_\star$, $R_\star$, and resulting planet properties). 

\begin{figure*}
  \includegraphics[width=12cm]{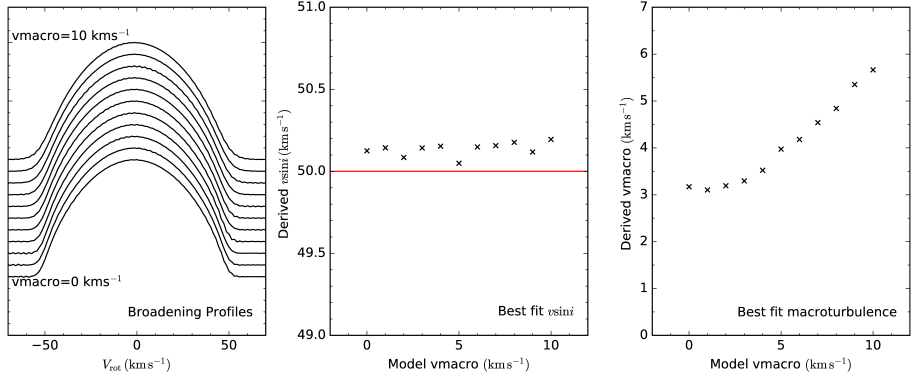}
  \caption{\label{fig:vmacro} We test the influence of macroturbulence (vmacro) on the $v\sin i$ derived from LSD broadening kernels. A series of synthetic spectra are generated, with $v\sin i=50\,\text{km\,s}^{-1}$, and macroturbulence of 0 to $10\,\text{km\,s}^{-1}$. Their derived broadening kernels are plotted on the left panel. The middle panel shows that $v\sin i$ can be accurately recovered if we account for both rotational and macroturbulent broadening in the LSD profile fitting. The recovered macroturbulence values are plotted on the left panel.}
\end{figure*}

To check if the $v\sin i$ can be accurately recovered from the combination of broadening factors, we generated a series of spectra with $v\sin i=50\,\text{km\,s}^{-1}$, macroturbulence of 0 to $10\,\text{km\,s}^{-1}$, and instrumental broadening of $6\,\text{km\,s}^{-1}$ (as per TRES resolution). These are deconvolved against an unbroadened template to derive broadening profiles for each test synthetic spectrum (Figure~\ref{fig:vmacro}). We fit the broadening kernel with the convolution of a rotation term \citep[modelled analytically from][]{2005oasp.book.....G} and a Gaussian term to account for macroturbulence \citep[expected for F-stars at $6500\,\text{K}$ to be $\sim 6\,\text{km\,s}^{-1}$][]{2014MNRAS.444.3592D} and instrumental broadening. This is different to the SPC approach, which cross correlates a series of spectral templates to the observed spectra, and maximises the cross correlation function peak. Figure~\ref{fig:vmacro} shows the $v\sin i$ can be recovered to within $0.2\,\text{km\,s}^{-1}$. From these tests, we also note that the $v\sin i$ measurement can be overestimated when we use a template that does not account for macroturbulence and instrumental broadening. 

A $v\sin i$ measurement is made for all available TRES out-of-transit spectra of each object. The median and standard deviation of the $v\sin i$ measurements are $69.2\pm0.2\,\text{km\,s}^{-1}$ and $35.7\pm0.7\,\text{km\,s}^{-1}$ for KELT-7 and HAT-P-56 respectively  We also derive macroturbulence velocities of $4.3\pm0.3\,\text{km\,s}^{-1}$ and $7.1\pm1.1\,\text{km\,s}^{-1}$ for each star respectively. The $v\sin i$ we derive for KELT-7 is consistent to that from the discovery paper to within errors ($65_{-5.9}^{+6.0}\,\text{km\,s}^{-1}$), while we derive a slower velocity for HAT-P-56 ($40.1\pm0.5\,\text{km\,s}^{-1}$ from discovery page), a difference likely attributed to the incorporation of macroturbulence in our analysis.   

\section{Global modelling of the transit geometry}
\label{sec:modelling}

To derive the spin-orbit angle of the planets, we performed a global modelling of the Doppler tomographic and photometric transit datasets. Since the parameters, such as transit depth, shape, and duration, are shared among the photometric and Doppler tomographic observations, a global fit is required to properly constrain the transit parameters and propagate associated uncertainties.

For KELT-7b, we included all the available photometric follow-up observations detailed in \citet{2015AJ....150...12B}. These included the University of Louisville Moore Observatory 2012-10-04 $g'$ band transit, FLWO KeplerCam 2012-10-23 $z'$, 2012-11-03 $z'$, 2013-11-22 $g'$, and 2013-10-19 $i'$ band transits, Bryne Observatory at Sedgwick 2012-11-14 $g'$ and 2014-01-13 $i'$ band transits, Canela's Robotic Observatory 2012-12-08 $V$, 2013-01-29 $i'$ band transits, and the Whitin Observatory at Wellesley College 2013-01-27 $i'$ band transit.

For HAT-P-56b, we included the K2 long cadence light curve available for the target. The K2 light curve reduction and detrending process are described in \citet{2015AJ....150...85H} and \citet{2015MNRAS.454.4159H}. 

The photometric transits were modelled as per \citet{2002ApJ...580L.171M}. Free parameters include the planet-star radius ratio $R_p/R_\star$, normalised orbital distance $a/R_\star$, orbital inclination inc, the transit centre time $T_0$, and period $P$. The quadratic limb darkening parameters for each band are taken from \citet{2011A&A...529A..75C}, interpolated using the tools described in \citet{2013PASP..125...83E} to the atmospheric parameters of each star, and fixed during the global fitting. To account for the long-cadence nature of the HAT-P-56 K2 light curves, we integrated the model over 30-minutes about each time stamp at 10 evenly spaced points. To ensure the per-point photometric uncertainties are accurate, we also inflated the uncertainties such that the reduced $\chi^2$ is at unity when compared against the best fitting model. 

\begin{figure}
  \begin{tabular}{c}
  \includegraphics[width=8cm]{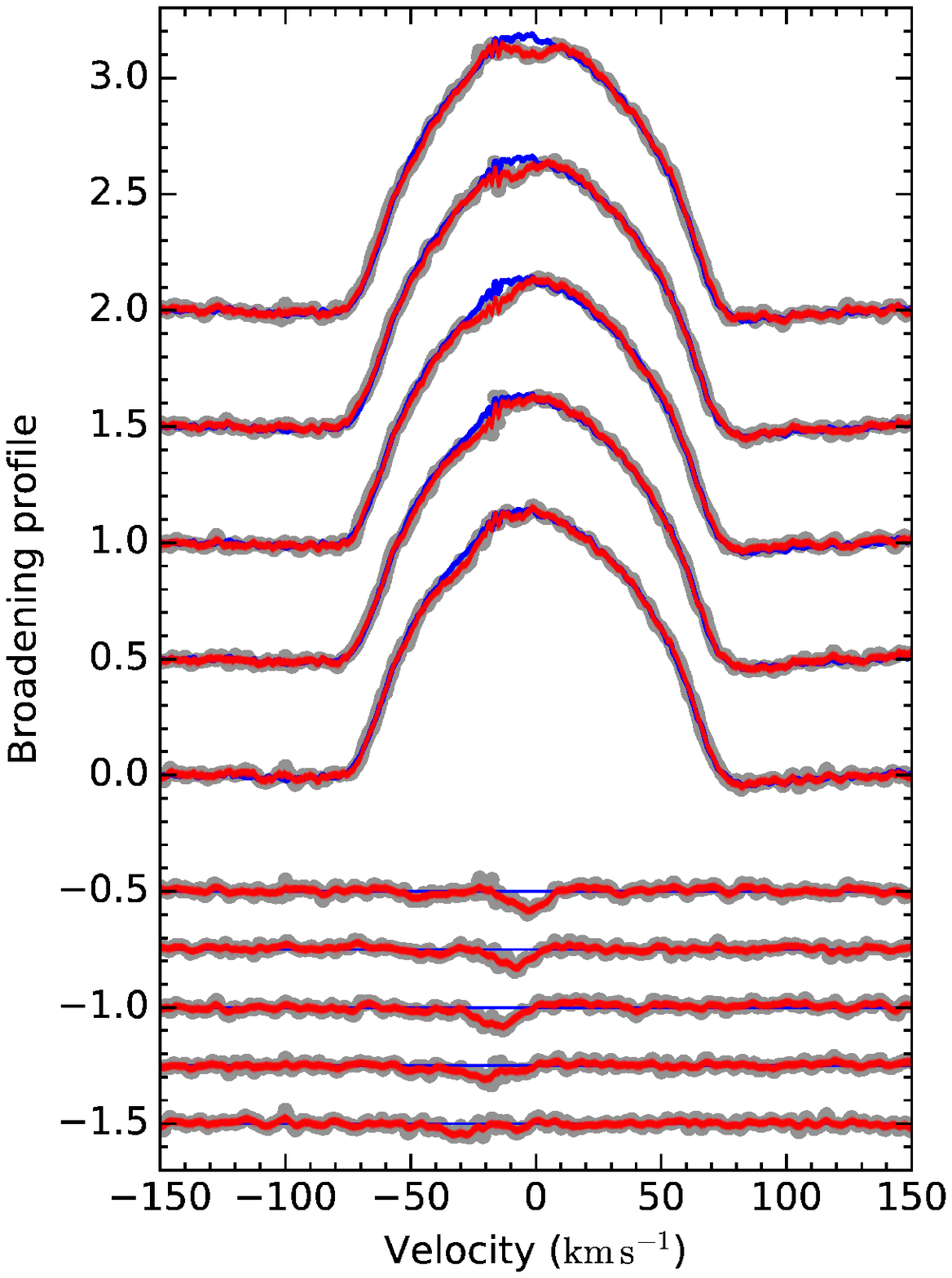}\\
  \includegraphics[width=8cm]{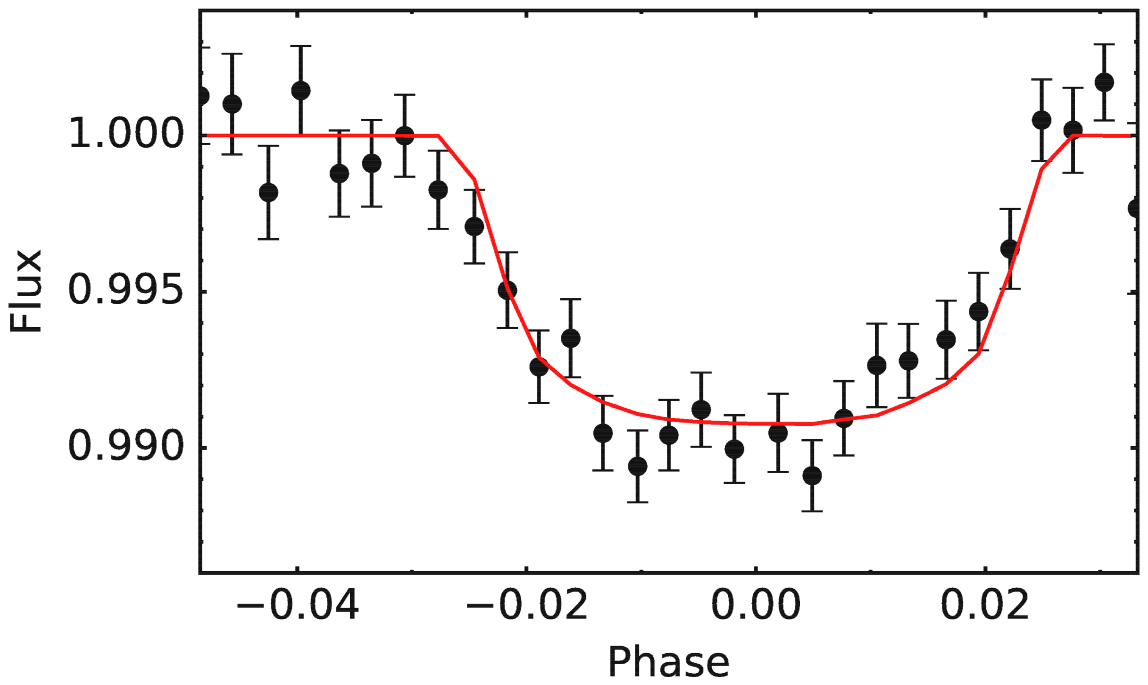}
  \end{tabular}
  \caption{\label{fig:vprof_KELT7}Top: Broadening kernels derived from a sample of five consecutive exposures during the transit of KELT-7b. The broadening kernel from each exposure is marked by the grey points. The average out-of-transit broadening kernel is marked by the blue lines. Models generated from a set of 100 randomly selected steps in the converged MCMC chain are plotted in red. The residual for each exposure from the average out-of-transit profile is shown below. The Doppler shadow of the planet can be seen in the residuals. Bottom: Measuring the area of the Doppler shadow is equivalent to measuring a transit light curve. The transit light curve of KELT-7b, directly measured from the Doppler tomographic signal, is plotted. }
\end{figure}

To model the Doppler tomographic transit observations, we first created an averaged out-of-transit rotational profile. The `shadow' of the planet is modelled as a Gaussian intrusion to the average rotational profile at each time step. The Gaussian has width of $R_p/R_\star \times v\sin i$, area of $1-f(t)$ (where $f(t)$ is the flux, blocked by the planet, that makes up the transit light curve), centred about $v_p(t)$ (where $v_p(t)$ is the projected rotational velocity for the region of the star occulted by the planet). Given the low S/N of the signal, a Gaussian function is a quick and effective model for the Doppler shadow of the planet \citep[e.g.][]{2016arXiv160200322C}. The parameter $v_p(t)$ is dependent on the spin-orbit angle $|\lambda|$, and the projected rotational velocity of the star $v \sin i$. As instrumental systematics can induce variations to the rotational profile at each time step, at each iteration we also fit for a dilation in the height and width of the rotational profile. An example of the rotational profile fitting, and the Doppler shadow of the planet, are shown in Figure~\ref{fig:vprof_KELT7} (Top). The flux blocked by the planet directly correlates with the area of the Doppler shadow, and we also constructed a transit light curve directly from the Doppler tomographic signal (Figure~\ref{fig:vprof_KELT7} Bottom).

The per point uncertainties in the rotational profile residuals were estimated by taking the standard deviation of the baseline regions of the rotational profile. We accounted for correlated noise in the broadening kernel via a Gaussian Processes approach \citep[e.g.][]{2015AJ....150..197H}. Applications of Gaussian process regression to astronomical signals have been extensively covered in the literature and shown to reliably retrieve model parameters from data-sets influenced by stochastic noise sources \citep[e.g.][]{2012MNRAS.419.2683G,2014MNRAS.445.3401G}. The Gaussian process regression was modelled with the \emph{George} module \citep{2014arXiv1403.6015A}. We employed a radial exponential kernel to model the co-variance $\Sigma_{ij}$ between points $i$, $j$, with velocities $v_{i}$ and $v_{j}$:
\begin{equation}
  \Sigma_{ij} = \sigma_{i}^2 \delta_{ij} + A \exp \left(- \frac{|v_{i}-v_{j}|}{\tau} \right)\,,
\end{equation}
where $\sigma_{i}$ is the per-point uncertainty for point $i$, and $\delta_{ij}$ is the Kronecker delta function. The Gaussian process hyper-parameters $A$ and $\tau$ specify the amplitude and the scale length of the covariance between the velocity points, respectively.

We explored the parameter space with a Markov chain Monte Carlo (MCMC) analysis, using the \emph{emcee} implementation \citep{2013PASP..125..306F} of an affine-invariant ensemble sampler. The transit parameters are also constrained by the spectroscopic stellar parameters for each star. At each iteration, we calculate an expected $a/R_\star$ using the orbital period, stellar mass and radii expected for the spectroscopic $T_\text{eff}$, $\log g$, [Fe/H] of the spectrum. The stellar mass and radius are interpolated from the spectroscopic parameters via the \citet{2010A&amp;ARv..18...67T} relationships. The expected $a/R_\star$ is then compared to the tested $a/R_\star$, thereby constraining the fit. The spectroscopic $v \sin i$ measurement and uncertainty from Section~\ref{sec:lsd} is applied to the fit as Gaussian prior. Gaussian priors were also imposed on the transit centre $T_0$ and period $P$, since these were derived in the discovery papers from the discovery and follow-up light curves, and are therefore much better constrained than from follow-up light curves alone. Uniform priors were assumed for all other parameters, including the hyper-parameters of the Gaussian process regression. 

\begin{figure*}
  \begin{tabular}{cc}
    \includegraphics[width=7cm]{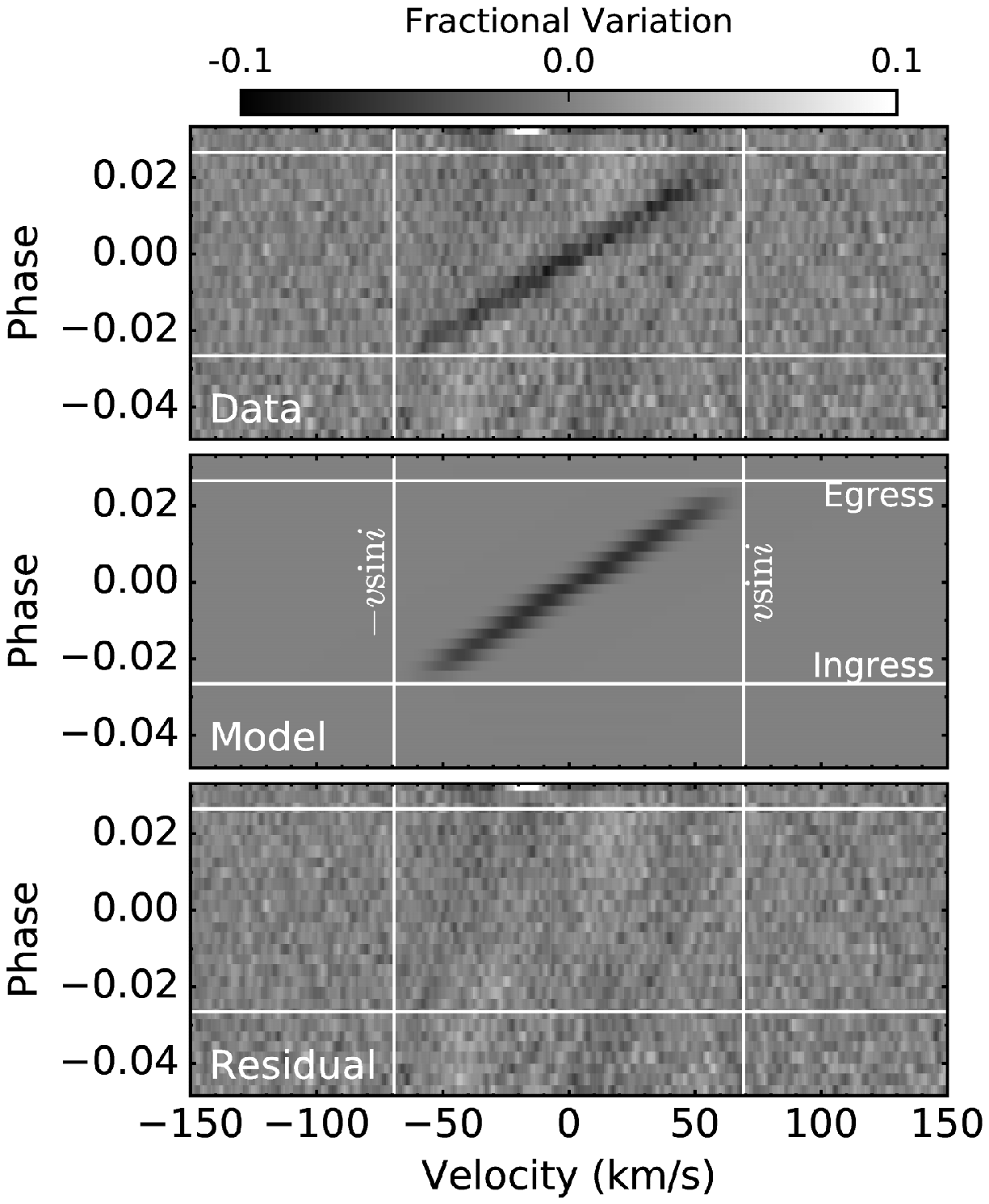} &
    \includegraphics[width=7cm]{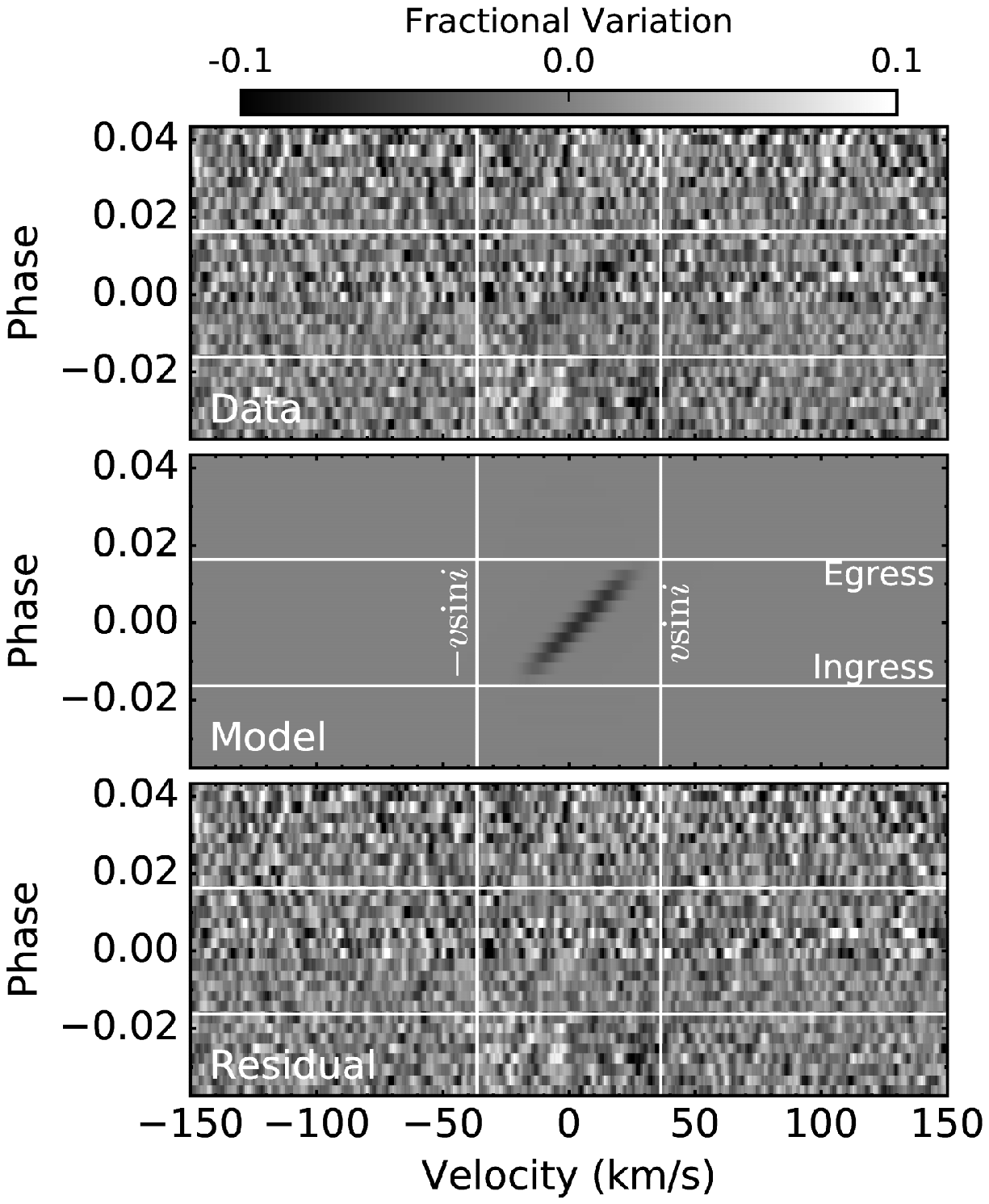} \\
  \end{tabular}
  
  \caption{\label{fig:dopplergram}The Doppler tomographic signal for KELT-7b (left) and HAT-P-56b (right) from the TRES observations. The top panels show the signal induced by the planet in the residual between each broadening kernel and the averaged out-of-transit kernel. The middle panels show the model of the best fit geometry. The bottom panels show the residual after the model is subtracted. }
\end{figure*}

The derived values and uncertainties are shown in Table~\ref{tab:parameters}. The Doppler tomographic signals, from the rotational profile fit to each TRES exposure, are shown in Figure~\ref{fig:dopplergram}. 

\begin{table*}
  
  \caption{\label{tab:parameters}Derived values for MCMC walker parameters}
  \begin{tabular}{lrr}
    \hline\hline
    & \textbf{KELT-7b} & \textbf{HAT-P-56b}\\
    \hline
    Period (days) $^a$ & $2.734780_{-0.000003}^{+0.000003}$& $2.790833_{-0.000004}^{+0.000004}$ \\
    $T_0$ (BJD)  $^a$  & $2456355.2293_{-0.0001}^{+0.0001}$ & $2456553.6164_{-0.0003}^{+0.0003}$ \\
    $R_p/R_\star$  & $0.0922_{-0.0004}^{+0.0004}$ & $0.099_{-0.002}^{+0.002}$ \\
    $a/R_\star$  & $5.50_{-0.06}^{+0.06}$ &  $6.7_{-0.4}^{+0.5}$ \\
    $\text{inc}\,(^\circ)$  & $83.7_{-0.2}^{+0.2}$ & $82.6_{-0.6}^{+0.7}$\\
    $|\lambda|\,(^\circ)$  & $2.7_{-0.6}^{+0.6}$ & $7_{-2}^{+2}$\\
    $v\sin i\,(\text{km\,s}^{-1})$  $^b$  & $69.3_{-0.2}^{+0.2}$ & $36.4_{-0.7}^{+0.7}$\\
    $\ln(A)$ $^c$  & $-9.17_{-0.04}^{+0.05}$ & $-7.65_{-0.05}^{+0.05}$ \\
    $\ln(\tau)$ $^c$ & $3.6_{-0.1}^{+0.1}$ & $2.24_{-0.1}^{+0.1}$\\
    $T_\mathrm{eff}$ (K) $^a$  & $6513_{-53}^{+49}$& $6568_{-53}^{+51}$\\
    $\log g$  & $4.14_{-0.03}^{+0.03}$ $^a$ & $4.26_{-0.05}^{+0.06}$\\
    \hline
  \end{tabular}
  \begin{flushleft}
  $^a$ Gaussian priors according to literature values were imposed.\\
  $^b$ Gaussian priors according to $v\sin i$ estimates from Section~\ref{sec:vsini} were imposed.  \\
  $^c$ Gaussian process hyper-parameters $A$ and $\tau$ describe the amplitude and the scale length of the covariance between velocity points in the broadening profile modelling.
  \end{flushleft}
\end{table*}

To test the dependence of our results on choice of Gaussian process co-variance kernel, we also tested an exponential squared kernel:
\begin{equation}
  \Sigma_{ij} = \sigma_{i}^2 \delta_{ij} + A \exp \left(- \frac{(v_{i}-v_{j})^2}{2\tau} \right)\,,
\end{equation}
yielding $|\lambda| = 2.6_{-0.6}^{+0.6}\,^\circ$ for KELT-7b and $|\lambda| = 7_{-2}^{+2}\,^\circ$ for HAT-P-56b.
Using the Matern 3/2 kernel:
\begin{equation}
  \Sigma_{ij} = \sigma_{i}^2 \delta_{ij} + A \left( 1+\sqrt{\frac{3(v_{i}-v_{j})^2}{\tau}} \right) \exp \left(-\sqrt{\frac{3 (v_{i}-v_{j})^2}{\tau}}\right)\,,
\end{equation}
yields $|\lambda| = 2.7_{-0.6}^{+0.6}\,^\circ$ for KELT-7b and $|\lambda| = 8_{-2}^{+2}\,^\circ$ for HAT-P-56b. In each case, the Gaussian process hyper-parameters converged to a solution without the need to apply priors.  To check for the effect of accounting for stochastic noise via Gaussian process, we also modelled the Doppler tomographic observations without allowing for co-variance between points, and derived similar results, but with smaller uncertainties, of $|\lambda|$ of $2.4_{-0.4}^{+0.4}\,^\circ$ for KELT-7b and $6_{+1}^{+2}\,^\circ$ for HAT-P-56b. 

To test the effect of a systematic offset in the assumed stellar parameters on our final result, we deconvolved the KELT-7 spectra against a 7000\,K template,  and modelled this new set of broadening profiles in our global MCMC analysis, whilst imposing a stellar parameter Gaussian prior of $T_\mathrm{eff}=7000$\,K, $\log g=4.2$ for the MCMC jump parameters. We arrive at the same set of planet parameters, with no significant change in the best fit values or uncertainties.

\section{Discussion}
\label{sec:discussion}

We find KELT-7b and HAT-P-56b to in spin-orbit aligned geometries, with $|\lambda|$ of $2.8 \pm 0.6\,^\circ$ and $8 \pm 2\,^\circ$ for the two systems respectively.

The transit parameters we derived for KELT-7b and HAT-P-56b are in general agreement with those in the discovery papers. KELT-7b is confirmed to be in a spin-orbit aligned geometry, in agreement with the Rossiter-McLaughlin analysis in \citet{2015AJ....150...12B}. In fact, \citet{2015AJ....150...12B} were able to detect the shadow of the planet in the cross correlation function of the spectra. Because of the grazing nature of the transit of HAT-P-56b, the uncertainties in its transit parameters are larger than the other systems. The uncertainties we derive for HAT-P-56b are are larger than those in the discovery paper, we find $R_p/R_\star = 0.099_{-0.002}^{+0.002}$ and $a/R_\star = 6.7_{-0.4}^{+0.5}$, compared to $R_p/R_\star = 0.1054\pm0.0009$ and $a/R_\star = 6.37\pm0.11$ from the discovery paper. The uncertainty in the planet radius is unchanged, since it is dominated by the uncertainty in the stellar radius, rather than in $R_p/R_\star$. We also note a weak stellar pulsation signal is seen in the Doppler tomographic analysis of KELT-7, manifested as diagonal stripes in Figure~\ref{fig:dopplergram}). Pulsations were not detected in the discovery KELT light curves of the star, but their presence not surprising given the star lies close to the instability strip. Similar pulsations are also seen in the Doppler tomographic observations of WASP-33b \citep{2010MNRAS.407..507C,2015ApJ...810L..23J} and HAT-P-57b \citep{2015AJ....150..197H}. 

\subsection{The spin-orbit angle distribution of hot-Jupiters around F-type stars}
\label{sec:stats}

A total of 6 hot-Jupiter systems around stars with $T_\text{eff} > 6250\,\text{K}$ have now been found in spin-orbit alignment $(|\lambda| < 10^\circ)$, while 15 have higher obliquities.

Have these systems undergone tidal synchronisation and realignment? The characteristic timescale for stellar spin synchronisation is given by \citet{2010ApJ...723..285H} and \citet{2012ApJ...757....6H}, who reviewed the tidal theory in the context of hot-Jupiter systems. Adopting Equation 3 in \citet{2012ApJ...757....6H}, the characteristic timescale to modify the spin of the host star, $T_\text{spin}$, is:
\begin{align}
  T_\text{spin} =& \frac{3.1\times10^9\,\text{years}}{(1-e^2)^{1/2}} \left(\frac{a}{0.02 \,\text{AU}} \right)^{7.5} \left( \frac{R_\star}{R_\odot}\right)^{-8} \left( \frac{k_0^2}{0.1} \right) \notag\\
  &\times \left( \frac{30 \,\text{days}}{P_\text{rot}} \right) \left(\frac{M_\star + M_p}{M_\odot} \right)^{1/2} \left(\frac{M_p}{M_\text{Jup}} \right)^{-2} \notag\\
  &\times \left( \frac{\sigma_\star}{7.8\times10^{-8}} \right)^{-1}\,.
\end{align}
We assume a stellar gyration radius of $k_0^2=0.1\,R_\star ^2$ \citep{2000MNRAS.315..543H}, stellar dissipation coefficients $\sigma_\star$ of $10^{-12}$ for KELT-7, and $10^{-8}$ for  HAT-P-56 \citep[from Figure 3 of][]{2012ApJ...757....6H}. The expected $T_\text{spin}$ is $\sim 10^{16}$ years for KELT-7 and $10^{12}$ years for HAT-P-56b. That is, tidal dissipation in the stellar envelope is expected to be weak for KELT-7 and HAT-P-56, at least based on standard equilibrium tide models. In fact, tidal interactions are expected to be weak for all the spin-orbit aligned hot-Jupiters around F-dwarfs. Only the transiting brown dwarf KELT-1b \citep{2012ApJ...761..123S}, found around a rapidly rotating F-star in a 1.2\,d period orbit, has a short tidal realignment timescale ($10^8$ years). 

KELT-7b and HAT-P-56b were found to be in projected spin-orbit alignment. If we assume alignment in the line-of-sight of the stellar spin axis as well, we can examine the spin-orbit coupling of these systems. The spins of KELT-7 and HAT-P-56 are both super-synchronous with respect to the orbital period of the planet. The maximum rotation period for KELT-7 is $1.08\pm0.03$ days, and $1.8\pm0.2$ days for HAT-P-56, based on the spectroscopic $v \sin i$ measurements. From the K2 light curves, \citet{2015AJ....150...85H} found that HAT-P-56 is a possible $\gamma$ Dor-pulsator with a primary pulsation period of $1.64\pm0.03$ days, and a secondary peak in the periodogram of $1.74\pm0.02$ days that is consistent with the $v\sin i$ derived rotation period. Neither the KELT discovery light curves, nor the archival SuperWASP light curves \citep{2010A&amp;A...520L..10B} yielded a photometric modulation period for KELT-7 in our analysis.  For comparison, \citet{2013MNRAS.436.1883W} found that \emph{Kepler} systems with planets of $R_p > 6 \,R_\text{E}$ and $\text{Period}<10$\,days are preferentially found in the stellar-spin -- planet-orbit synchronised states. The four that were systems found to be in super-synchronous states had orbital periods greater than 5 days, the largest of which had a radius of $0.7\,R_\mathrm{J}$. For a consistency check on the assumption that these two systems are also in line-of-sight alignment, we can compare the $v\sin i$ of these stars to that expected from the rotation periods of Kepler stars of similar stellar parameters \citep{2013A&amp;A...557L..10N}. For F-dwarfs like KELT-7 ($6600 < T_\mathrm{eff} < 6800$), 68\% of stars have rotation periods that lie within 1.2 -- 5.9 days. The $v\sin i$-derived rotation period of the KELT-7 is 1.08 days, consistent with the population, and with an aligned geometry. For stars like HAT-P-56 ($6400 < T_\mathrm{eff} < 6600$), 68\% of stars lie within rotation periods of 1.4 -- 8.6 days. The rotation period of HAT-P-56 from $v\sin i$ is 1.8 days, again consistent with the distribution, and with alignment.

\begin{figure*}
  \includegraphics[width=14cm]{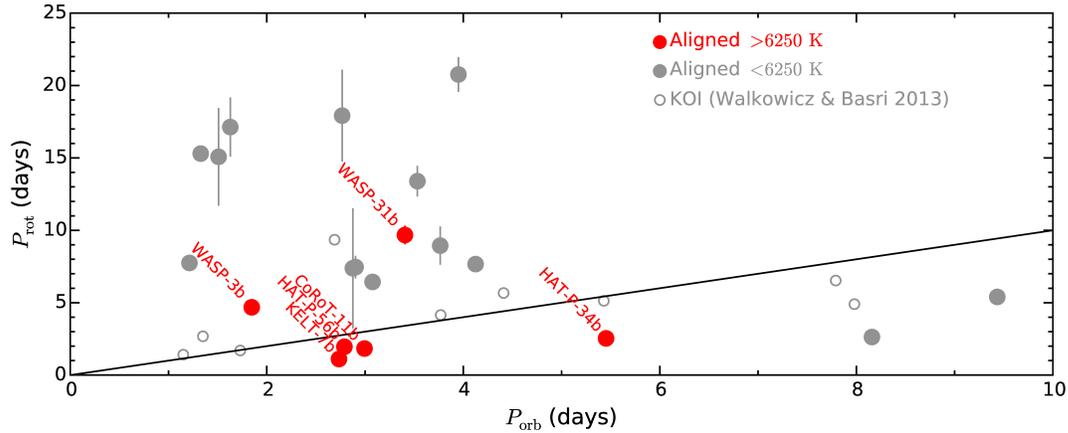}
  \caption{\label{fig:porb_prot}The orbital periods $P_\text{orb}$ and stellar rotation periods $P_\text{rot}$ of close-in hot-Jupiter systems in spin-orbit alignment $(|\lambda| < 10^\circ)$. The solid line marks 1:1 spin-orbit synchronisation. KELT-7 and HAT-P-56 are super-synchronous with respect to the orbital period of their planets. The rotation of Kepler candidates were derived by \citet{2013MNRAS.436.1883W} from their light curves. Otherwise, the spin rotation periods are inferred from the spectroscopic $v\sin i$, assuming $i=90^\circ$, and as such represent the upper limit of the rotational periods. Super-synchronous rotation for short period systems ($P<5$\,days) are only found in systems with host stars of $T_\text{eff}>6250\,\text{K}$. We note that one of the $|\lambda|$ solutions for HAT-P-57b is of low obliquity and super-synchronous, but given the ambiguity that multiple $|\lambda|$ values are allowed \citep{2015AJ....150..197H}, it is left off the plot.}
\end{figure*}

Along with the CoRoT-11 system \citep{2010A&amp;A...524A..55G,2012A&amp;A...543L...5G}, KELT-7 and HAT-P-56 are the only spin-orbit aligned super-synchronous systems with planetary orbital periods $<5$\,days. Figure~\ref{fig:porb_prot} shows the orbital period $P_\text{orb}$ of spin-orbit aligned systems against their stellar rotation period $P_\text{rot}$. With the exception of the Kepler candidates from \citet{2013MNRAS.436.1883W}, the $P_\text{rot}$ values are derived from the spectroscopic $v\sin i$ measurements, which should be representative of the stellar spin period if we assume these systems are truly aligned. The only close-in systems in super-synchronous states are found around F-dwarfs, but this may be a selection bias due to the lack of rapidly rotating, cooler stars. For these aligned super-synchronous systems, the angular momentum exchange between the star and the planet is expected to slow down the rotation of the star, and extend the orbital period of the planet. However, the timescale for tides to modify the orbital period is similar to that of the stellar spin synchronisation timescales \citep{2012ApJ...757....6H} for KELT-7 and HAT-P-56, and should not have affected the orbital periods of the planets. We note that a number of other spin-orbit misaligned systems are also found in super-synchronous states (CoRoT-3b, KOI-13b, WASP-7b, WASP-8b, WASP-33b, WASP-38b). 

\begin{figure}
  \includegraphics[width=8cm]{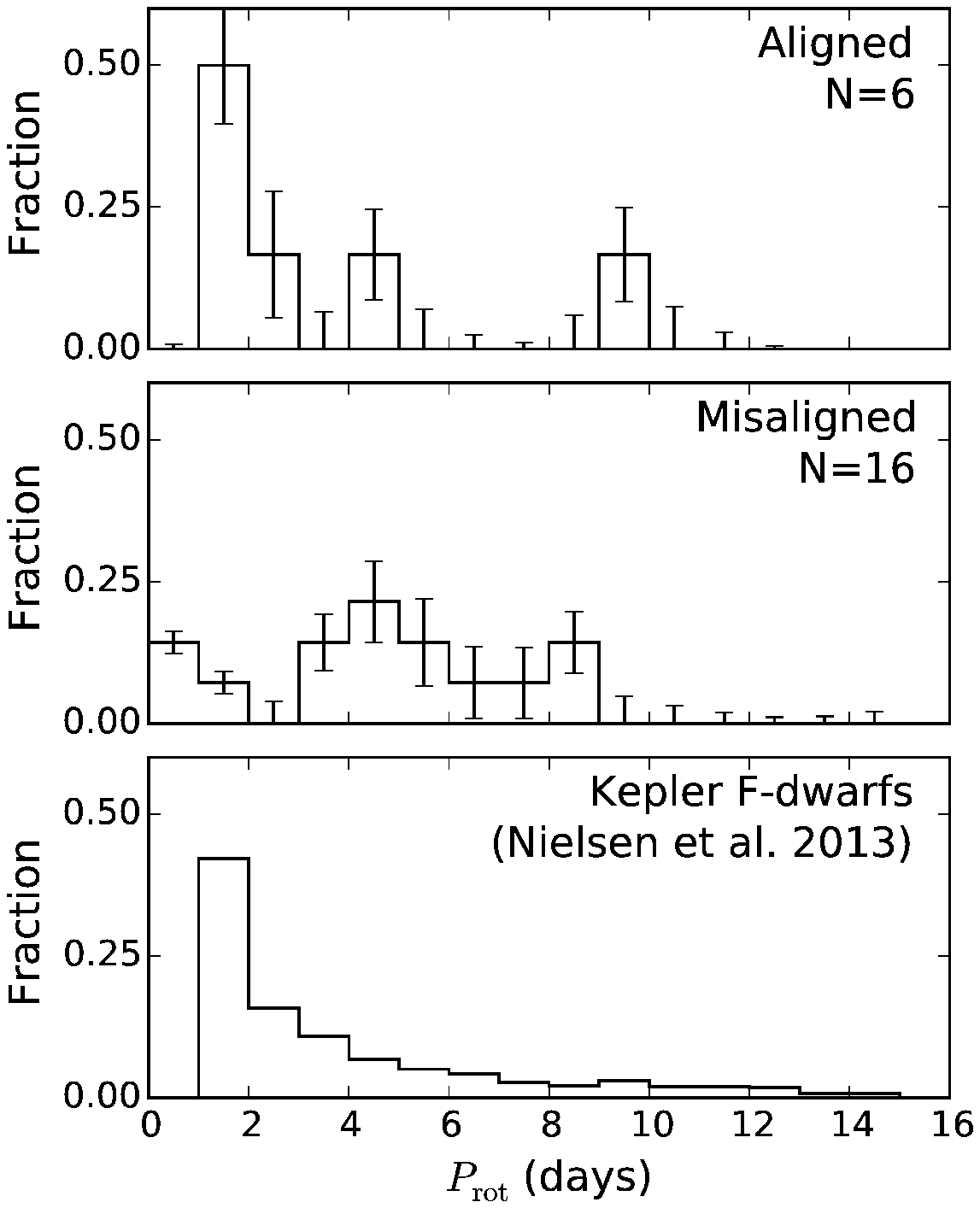}
  
  \caption{\label{fig:prot_hist}The distribution of stellar rotation periods for systems orbiting hot stars $(T_\text{eff}>6250\,\text{K})$. The rotation periods are inferred from their spectroscopic $v\sin i$, assuming $i=90^\circ$. The top panel shows the distribution for well-aligned systems, middle panel shows systems with misalignment in $|\lambda|$. The bottom panel shows the distribution for the rotation periods for stars of the equivalent spectral type measured from Kepler photometry by \citet{2013A&amp;A...557L..10N}}
\end{figure}

To further check for tidal evolution of the stellar spin, we can also compare the derived rotation periods of F-stars hosting large transiting planets against a similar sample without transiting hot-Jupiters. Figure~\ref{fig:prot_hist} shows the distribution of rotation periods for host stars with $T_\text{eff}>6250\,\text{K}$, binned into aligned $(|\lambda| < 10^\circ)$ and misaligned groups. We also show the rotation period distribution for equivalent stars from the Kepler sample \citep{2013A&amp;A...557L..10N}. To check for the distinction between the populations, we run a two-sample Kolmogorov-Smirnov over 100000 iterations. At each iteration, we draw samples from the rotation period distribution via a bootstrap process, and then draw the rotation period of each star from a Gaussian distribution with the standard deviation as their respective uncertainties. The rotation period of the Kepler F-dwarf sample and that of F-dwarfs hosting spin-orbit aligned systems cannot be distinguished $(p=0.6\pm0.2)$. However, we can tentatively reject the null hypothesis that the rotation periods of the misaligned sample originated from the same population as the Kepler F-dwarfs $(p=0.017\pm0.019)$, although it is marred by small-number statistics. Similarly, a K-sample Anderson-Darling test cannot distinguish between the non-transit planet hosting F-dwarf sample and the spin-orbit aligned sample ($p=0.7\pm0.2$), but can distinguish against the misaligned sample ($p=0.009\pm0.008$). This is expected given these systems are already known to have misaligned $|\lambda|$ angles, and are likely to have line-of-sight $(i)$ misalignments too. The uncertainties are derived by a Monte Carlo exercise, drawing the rotation period of each star from a Gaussian distribution with the standard deviation as their respective uncertainties. This suggests that 1) there is no evidence that the rotation periods of hot stars hosting spin-orbit aligned planets have being modified by tidal interactions, 2) the systems with low projected obliquities are likely to have low true obliquities too. We note that this analysis suffers from an observational selection bias against rapidly rotating stars. Planets found against rapid rotators are difficult to confirm, and therefore lacking in the literature. The same analysis on planets orbiting cool stars $(T_\text{eff}<6250\,\text{K})$ could not distinguish between any of the populations, since these cools stars are spun-down with age, and do not exhibit a sharp rotation period distribution.

Given the lack of evidence for tidal evolution in the rotation periods of most hot host stars, we can examine the set of spin-orbit angles for these systems around hot stars in the context of migration mechanisms. 23\% of the systems around hot stars are found in spin-orbit aligned arrangements. While the fraction of aligned systems is significantly lower than that of the overall distribution, it is still different from the relatively even $\lambda$ distribution expected from dynamical interactions such as eccentric migration via stellar binary Kozai-Lidov cycles \citep[e.g.][]{2012ApJ...754L..36N,2015ApJ...799...27P} or planet-planet scattering \citep[e.g.][]{2011ApJ...742...72N}. Nevertheless, dynamical interactions, compared to in-disk co-planar migration, are likely responsible for a significant fraction of hot-Jupiters around hot stars. It should be noted that inhomogeneity of the star-forming cloud, or binary-induced disk tilting, will also cause primordial spin-orbit misalignment \citep{2010MNRAS.401.1505B,2012Natur.491..418B}. We also note \citet{2012ApJ...758L...6R} suggest that internal gravity waves at the convective core -- radiative envelope boundary can induce arbitrary surface spins for hot stars, independent of star-planet interactions.

\section*{Acknowledgements}
\label{sec:acknowledgements}
GZ thanks Chelsea Huang for HAT-P-56b K2 light curves. The modelling in this paper was performed on the Smithsonian Institution High Performance Cluster (SI/HPC). We also thank Jessica Mink for running the TRES pipeline and maintaining the TRES archive. We acknowledge Andrew H. Szentgyorgyi, Gabor F\H{u}r\'{e}sz, and John Geary, who played major roles in the development of the TRES instrument.

\bibliographystyle{mn2e}

\bibliography{mybibfile.bib}

%\FloatBarrier

\label{lastpage}

\end{document}